\documentclass[aps,prl,superscriptaddress,longbibliography,nobibnotes,
twocolumn
]{revtex4-2}

\usepackage[T1]{fontenc}
\usepackage[utf8]{inputenc}
\usepackage{amsmath}
\usepackage{amssymb}
\usepackage{bm}
\usepackage{graphicx}
\usepackage{hyperref}
\usepackage{xcolor}

\graphicspath{{figs/}}

\begin{document}

\title{Residual orbital magnetization governs the anomalous Hall effect in altermagnets}

\author{Yufei Zhao}
\affiliation{Department of Condensed Matter Physics, Weizmann Institute of Science, Rehovot 7610001, Israel}
\affiliation{Department of Physics, The Pennsylvania State University, University Park, Pennsylvania 16802, USA}
\author{Yiyang Jiang}
\affiliation{Department of Physics, The Pennsylvania State University, University Park, Pennsylvania 16802, USA}
\author{Kamal Das}
\affiliation{Department of Condensed Matter Physics, Weizmann Institute of Science, Rehovot 7610001, Israel}
\affiliation{Department of Physics, The Pennsylvania State University, University Park, Pennsylvania 16802, USA}
\author{Chao-Xing Liu}
\affiliation{Department of Physics, The Pennsylvania State University, University Park, Pennsylvania 16802, USA}
\affiliation{Center for Theory of Emergent Quantum Matter, The Pennsylvania State University, University Park, Pennsylvania 16802, USA}
\author{Binghai Yan}
\email{binghai.yan@psu.edu}
\affiliation{Department of Condensed Matter Physics, Weizmann Institute of Science, Rehovot 7610001, Israel}
\affiliation{Department of Physics, The Pennsylvania State University, University Park, Pennsylvania 16802, USA}
\affiliation{Center for Theory of Emergent Quantum Matter, The Pennsylvania State University, University Park, Pennsylvania 16802, USA}

\date{\today}

\begin{abstract}
\textbf{Abstract:} In altermagnets that exhibit anomalous Hall effect, the small remanent magnetization exists but has been treated as too small to be relevant to the Hall response. 
In this work, we point out that this dismissal is incomplete because the generalized St\v{r}eda relation ties the intrinsic anomalous Hall conductivity ($\sigma_{xy}$) to the orbital magnetization ($M_z$, the topological component from the modern orbital magnetization) by $\sigma_{xy}=-e\frac{\partial M_z}{\partial \mu}$. We reveal a microscopic mechanism to generate net orbital moment from the interplay of local crystal field and spin-orbit coupling for MnTe-type altermagnets, in which the magnetic anisotropy generates weak net magnetization without invoking exchange between neighboring spins (e.g., Dzyaloshinskii-Moriya interaction). Our work indicates that residual orbital and spin magnetization is an intrinsic thermodynamic property that governs anomalous transport in unconventional antiferromagnets, including altermagnets and noncollinear antiferromagnets. 
\end{abstract}

\maketitle

\section{Introduction}
Altermagnets are a recently identified class of collinear compensated magnets that combine vanishing net magnetization with momentum-dependent spin splitting and can support the anomalous Hall effect (AHE) \cite{Smejkal2022,Smejkal2022Landscape,MazinEditorial2022,Mazin2023,Liu2025NP,Song2025NRM,Wu2007,Yuan2020,Ma2021,liu2026symmetry,jungwirth2025altermagnetism}.  Representative members such as $\alpha$-MnTe exhibit a clear anomalous Hall response together with a remanent magnetization ($\sim 10^{-5} ~\mu_B$ per Mn atom) \cite{GonzalezBetancourt2023,Kluczyk2024}. 
This echoes the older non-collinear antiferromagnet family Mn$_3$Sn and Mn$_3$Ge in which a sizable AHE was observed with a tiny net magnetization ~\cite{Nakatsuji2015,Nayak2016,Chen2014anomalous,Zhang2017strong}.  
Across both classes the remanent magnetization is small enough that it has been routinely treated as too tiny to be relevant to the Hall response according to the empirical formula~\cite{Nagaosa2010anomalous}, 
\begin{equation} \label{eq:AHErho}
\rho_H = R_0H + R_sM.   
\end{equation}

Thus far, the remanent magnetization is rarely appreciated in literature. $\alpha$-MnTe provides a representative case in which this dismissal pattern is explicit. Experiments report a small but reproducible remanent moment alongside a clear AHC \cite{GonzalezBetancourt2023,Kluczyk2024}. The prevailing interpretations of this moment fall into three classes. The first treats it as a materials-specific or extrinsic background tied to stoichiometry and sample conditions.\cite{Kluczyk2024} The second keeps the origin intrinsic but views the moment as a higher-order spin-canting residue induced by spin-orbit coupling~\cite{MazinBelashchenko2024,Jo2025} or Dzyaloshinskii-Moriya interaction~\cite{autieri2025stagger}, again secondary to the Hall mechanism, merely admitting the symmetry-allowed coexistence of residual magnetization and AHE~\cite{smejkal2022anomalous,McClarty2024}.
The third, identifies the orbital component as the nonnegligible contribution to the net moment~\cite{Chen2020,Ye2025MnTeOrbital} but does not yet promote it to the bulk observable that carries the Hall response. In each reading the small remanent magnetization is treated as either an accident or a perturbative residue, leaving the underlying question untouched. Except MnTe~\cite{liu2026observation}, the weak ferromagnetization was reported together with AHE in many other altermagnet materials such as FeS ($10^{-4}~\mu_B$)~\cite{takagi2025spontaneous} and $\mathrm{Mn_5Si_3}$ ($10^{-2}~\mu_B$)~\cite{badura2025observation}.

We point out that the orbital magnetization among the remanent net magnetization governs the intrinsic AHE in these unconventional antiferromagnets. Berry-phase modern theory treats the orbital magnetization ($M_z$) as a bulk quantity and links its chemical-potential derivative to the anomalous Hall conductivity (AHC,$\sigma_{xy}$) through the generalized St\v{r}eda relation,\cite{Xiao2010berry}
\begin{equation}
\sigma_{xy} = -e \left(\frac{\partial n}{\partial B}\right)_{\mu,T} =
-e \left(\frac{\partial M_z}{\partial \mu}\right)_{B,T},
\label{eq:streda}
\end{equation}
which generally applies to insulators, for example, the case of altermagnetic quantum anomalous Hall insulator~\cite{jiang2026altermagnetism}. In metals, above St\v{r}eda relation for the \textit{intrinsic} AHC involves only the Berry curvature component of the orbital magnetization~\cite{Xiao2005,Ceresoli2006,Shi2007,Souza2008} despite that the total orbital magnetization still contains a trivial part. 
So the decisive quantity is the derivative of orbital magnetization to the chemical potential ($\mu$) but not its magnitude. A simple sensitivity estimate already shows why the small-moment objection is too quick: 
If an orbital moment of $10^{-5}~\mu_B$ varies slightly, e.g., by $1\%$ over $1$ meV, it corresponds to $\sigma_{xy} \approx 0.1$ S/cm (assume the unit cell volume about 100 \AA$^3$) on the experimentally relevant scale (e.g., $0.01\sim0.1$ S/cm for MnTe in  Ref. \cite{GonzalezBetancourt2023}).

In this work, we demonstrate that the orbital magnetization governs the intrinsic AHE via the St\v{r}eda relation (Eq.~\ref{eq:streda}) in the altermagnet by using an effective model. Here, AHE persists even when the orbital magnetization accidentally reaches zero. Furthermore, we show the emergence of out-of-plane orbital and spin moments in a local octahedral model with nonlinear dependence on the spin-orbit coupling (SOC). The unique crystal field - spin sublattice locking leads to the same spin canting effect among two spin sublattices, generating a net magnetization. These residual moments form weak ferromagnetic order coexisting with in-plane Néel order, consistent with experimental observations. Our work clarifies the vital role of residual magnetization in unconventional antiferromagnetism, which is so far largely overlooked. 

\section{Results}
\subsection{St\v{r}eda relation}
The orbital magnetization has the modern formulation~\cite{Xiao2005,Ceresoli2006,Shi2007}, 
\begin{equation}
M_z(\mu)
   = M_z^{\rm tri}(\mu) + M_z^{\rm topo}(\mu),
\label{eq:Mz}
\end{equation}
where the two components are
\begin{align}
M_z^{\rm tri}(\mu)
   &= \frac{e}{\hbar}\sum_n\!\int\!\frac{d^3k}{(2\pi)^3}\;
      f_n(\bm k)\,
      \mathrm{Im}\!\!\sum_{m\neq n}\!
      \frac{v^x_{nm}\,v^y_{mn}}
           {\varepsilon_m - \varepsilon_n},
   \label{eq:Mz_tri}\\[2pt]
M_z^{\rm topo}(\mu)
   &= -\,\frac{e}{\hbar}\sum_n\!\int\!\frac{d^3k}{(2\pi)^3}\;
      f_n(\bm k)\,\bigl(\varepsilon_n(\bm k) - \mu\bigr)\,
      \Omega_n^{xy}(\bm k).
   \label{eq:Mz_topo}
\end{align}
The first contribution, $M_z^{\rm tri}$, is the conventional ``trivial''
part due to wave-packet self-rotation, while the second, $M_z^{\rm topo}$, is the topological part from the Berry curvature related to the mass center motion. In Eqs.~(\ref{eq:Mz_tri})--(\ref{eq:Mz_topo}), $M_z$ is an orbital
magnetization \emph{density} (moment per unit volume), as fixed by the
$\int d^3k/(2\pi)^3$ normalization, so that the St\v{r}eda relation
Eq.~(\ref{eq:streda}) 
is dimensionally a three-dimensional conductivity.
For readability, the figures and quoted values express it as a moment per
unit cell, $V_{\rm cell}M_z$, with $V_{\rm cell}\approx 100~\text{\AA}^3$
for $\alpha$-MnTe (two Mn atoms per cell); experimental remanent moments are
quoted per Mn atom, $V_{\rm cell}M_z/2$.

The intrinsic anomalous Hall conductivity is calculated in the Kubo--Berry-curvature form
\begin{equation}
\sigma_{xy}^{\rm AH}(\mu)
   = -\frac{e^{2}}{\hbar}\sum_{n}\!\int\!\frac{d^{3}k}{(2\pi)^{3}}\,
     f_{n}(\bm k)\,\Omega_{n}^{xy}(\bm k),
\label{eq:sigma}
\end{equation}
with $\Omega_{n}^{xy}(\bm k)= -2\sum_{m\neq n}\mathrm{Im}[v^{x}_{nm}v^{y}_{mn}]/[\varepsilon_{n}-\varepsilon_{m}]^{2}$. In the following we are going to demonstrate the zero-temperature St\v{r}eda relation between $\sigma_{xy}^{\rm AH}$ calculated by Eq.~\ref{eq:sigma} and $M_z^{\rm topo}$ by Eq.~\ref{eq:Mz_topo},
\begin{equation}
\sigma_{xy}^{\rm AH} = - e \frac{\partial M_z^{\rm topo}}{\partial \mu}.
\label{eq:streda2}
\end{equation}

Without loss of generality, we evaluate the orbital magnetization and AHE using a four-band model for $g$-wave altermagnet on the hexagonal lattice (point group $D_{6h}$) introduced in Ref.~\cite{Roig2024PRB}. Magnetic atoms form a hexagonal lattice (Wyckoff $2a$) with site symmetry $D_{3d}$. The two sublattices $A,B$ of the magnetic ions in the unit cell will be labeled by the Pauli matrices $\bm \tau$, and the spin degree of freedom is labeled by the Pauli matrices $\bm \sigma$. The effective Hamiltonian is under the basis order of $\{|A\!\uparrow\rangle,|A\!\downarrow\rangle,|B\!\uparrow\rangle,
|B\!\downarrow\rangle\}$,  written as
\begin{equation}
H_{\bm k} = \varepsilon_{0,\bm k}\,\tau_0 \sigma_0
            + t_{x,\bm k}\,\tau_x \sigma_0
            + t_{z,\bm k}\,\tau_z \sigma_0
            + \tau_y \bm\lambda_{\bm k}\!\cdot\!\bm\sigma
            + \tau_z \bm J\!\cdot\!\bm\sigma.
\label{eq:H}
\end{equation}
Here, $t_{x,\bm k}$ represents the $k_z$ dispersion, $t_{z,\bm k}$ includes the anisotropic dispersion, $\tau_y \bm\lambda_{\bm k}\!\cdot\!\bm\sigma $ is the SOC interaction and $\tau_z \bm J\!\cdot\!\bm\sigma$ is the staggered Néel exchange, where two sublattices exhibit opposite in-plane spins. See details of model parameters in the Methods and Supplementary Information (SI).

\begin{figure*}
    \centering
    \includegraphics[width=1\linewidth]{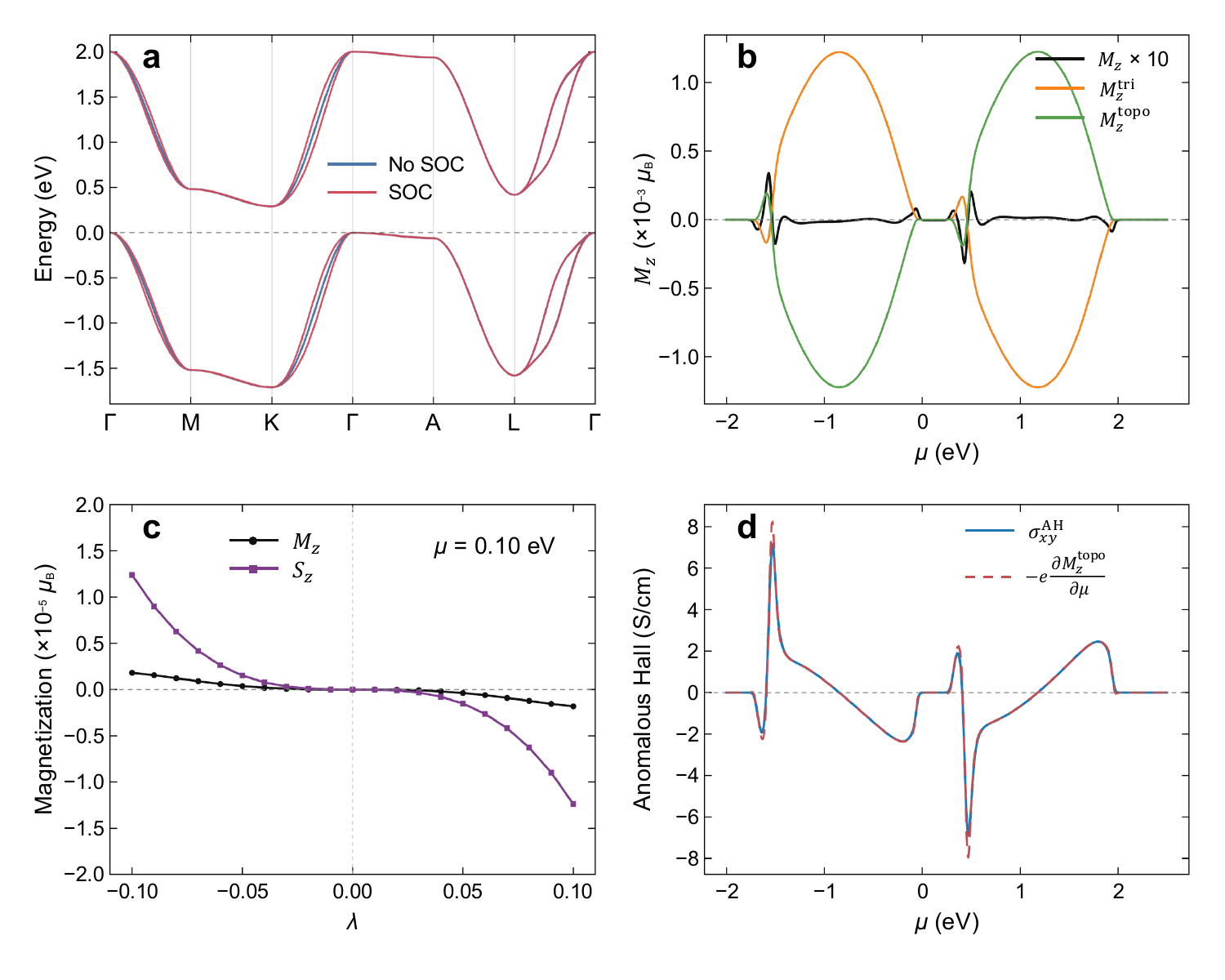}
    \caption{Band-structure and response summary of the four-band $g$-wave altermagnet model. (a) Band dispersion along the high-symmetry path with (red) and without (blue) spin-orbit coupling (SOC). The parameters are chosen as $t_1$ = 0.38, $t_2$ = 0.03, $t_3,t_4,\lambda$ = 0.05, and $J_y$ = 1.00 (all in eV). (b) Orbital magnetization decomposition as a function of chemical potential $\mu$: total $M_z$ (black line, multiplied by 10 for visibility), trivial contribution $M_z^{\mathrm{tri}}$ (orange) and topological contribution $M_z^{\mathrm{topo}}$ (green). (c) Orbital magnetization $M_z$ (black) and spin magnetization $S_z$ (purple) as a function of SOC strength ($\lambda$) in the gap ($\mu = 0.10$ eV). (d) Anomalous Hall conductivity $\sigma_{xy}^{\mathrm{AH}}$ (blue) compared with the St\v{r}eda proxy $-e \frac{\partial M_z^{\mathrm{topo}}}{\partial \mu}$ (dashed red line), illustrating the equivalence of two quantities.}
    \label{fig:band}
\end{figure*}

We consider an in-plane spin orientation with $\bm J=(0, J, 0)$, which mimics the in-plane spin order of MnTe. The band structure is shown in Fig.~\ref{fig:band}(a). The four bands group into two N\'eel-split doublets separated by $\sim 2J$. The non-SOC spin split along $\Gamma-L$ indicates the altermagnetic order. Then SOC weakly gaps the degeneracy along $\Gamma-M$ and $\Gamma-K$. These SOC-induced gaps are sources of Berry curvature and generate the AHE and orbital magnetization. 

Figure ~\ref{fig:band}b shows a small but finite orbital moment $M_z \sim 10^{-5}~\mu_B$ per unit cell while its $M_z^{\rm tri}$ and $M_z^{\rm topo}$ components are much larger ($\sim 10^{-3}~\mu_B$) in magnitude than $M_z$. 
We can extract the AHC from the St\v{r}eda formula (Eq.~\ref{eq:streda2}) and find consistence with $\sigma_{xy}^{\rm AH}$ directly calculated from Eq.~\ref{eq:sigma}, as shown in Fig.\ref{fig:band}(d). 
Inside the energy gap, $M_z$, $M_z^{\rm tri}$ and $M_z^{\rm topo}$ are constant and thus $\sigma_{xy}^{\rm AH}=0$. Near conduction band edges (e.g. $\mu \approx 0.29$ eV), $M_z$ or $M_z^{\rm topo}$ may accidentally turn zero while $\sigma_{xy}^{\rm AH}$ does not vanish but exhibits peaks in these regions. In some chemical potential window (e.g., $\mu \approx -0.84$ or 1.17 eV),  $\sigma_{xy}^{\rm AH}$ can changes sign while the residual magnetization remains the same sign, in which AHE cannot be explained merely by the magnetization. 
It is clear that the orbital magnetization magnitude does not directly determine the AHE, contradicting the empirical relation in Eq.~\ref{eq:AHErho}.

The orbital magnetization comes together with an out-of-plane spin polarization ($S_z$), which indicates weak out-of-plane spin canting of Mn moments. As shown in Fig.~\ref{fig:band}(c), $S_z$ originates from SOC with a third-order nonlinear dependence, consistent with Ref.~\cite{MazinBelashchenko2024}. 
This weak out-of-plane ferromagnetism is consistent with the magnetic susceptibility\cite{Kluczyk2024} and neutron scattering\cite{liu2026observation}, and magnetic imaging \cite{zhou2026imaging} in MnTe. 
It is clear that $M_z$ and $S_z$ are definite in magnitude and sign for fixed spin order and SOC. When rotating  $\bm J$ in the $xy$ plane, we find a $\sin(3\phi)$-type angle dependence for $M_z$ and $S_z$ (we choose $\phi=0^\circ$ along the $x$ axis, see details in SI), consistent with previous AHE calculations~\cite{Smejkal2020CHE,Mazin2023}. Thus, signs of $M_z$ and $S_z$ reverse if the in-plane spins flip, to reduce the ground state energy.

\subsection{Altermagnet molecule}
So far, the out-of-plane magnetization in altermagnets has been accessed primarily from a momentum-space perspective, through the integration of Berry curvature over the Brillouin zone, as performed above. A complementary real-space understanding based on the local atomic structure is still lacking. Existing explanations often invoke exchange-mediated canting of neighboring magnetic moments, such as that arising from Dzyaloshinskii–Moriya or similar interactions \cite{MazinBelashchenko2024,autieri2025stagger,zhou2025manipulation,Khodas2026}. An outstanding question is therefore whether the spin canting is intrinsically a collective phenomenon requiring intersite exchange, or whether it can already emerge from spin–orbit coupling within a single local atomic environment.

\begin{figure*}
    \centering
    \includegraphics[width=0.8\linewidth]{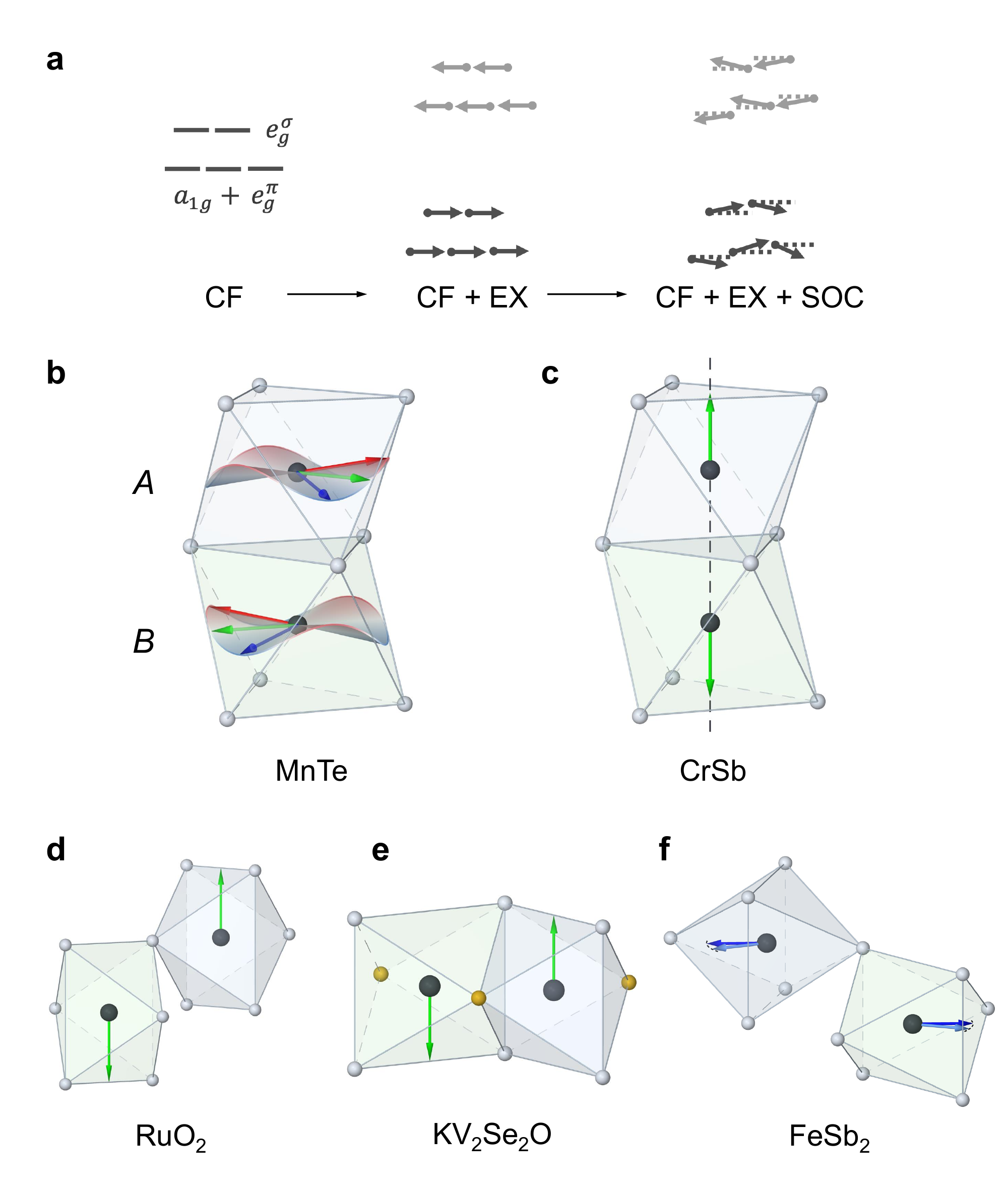}
\caption{Altermagnet molecule model: microscopic origin of spin canting in the magnetic anisotropy of the local octahedron. (a) Evolution of the Mn-$3d$ energy levels (short horizontal bars) and spin states by sequentially including three terms to the local Hamiltonian (Eq.~\ref{eq:Hloc}), the crystal field (CF), exchange interaction (EX), and spin-orbit coupling (SOC), for the in-plane spin ($\phi=30^\circ$). SOC induces spin canting by mixing two spin channels. Arrows indicate spin directions and the out-of-plane tilt is exaggerated for clarity. (b)--(f) Local octahedra for $A$/$B$ sublattices and spin orientations in altermagnet candidates. Black (white) spheres represent the cation (anion) atom, and light-blue and light-green octahedra belong to the sublattices $A$ and $B$, respectively. (b) MnTe: Angle($\phi$)-dependent spin canting in the $\mathrm{MnTe_6}$ octahedra. The shaded ribbon around each Mn atom traces the out-of-plane spin tilt as $\phi$ varies. Red (blue) arrow stands for canting up (down) while the green arrow stands for no canting ($\phi=0^\circ$). (c) CrSb: the spin (green arrows) points along the $z$ axis (dashed line) toward the center of an octahedral face, an equilateral triangle, so no canting direction is preferred. (d,e) RuO$_2$ and KV$_2$Se$_2$O: the spin points toward the midpoint of an octahedral edge with no preferred canting direction. Gold spheres in (e) are O anions. (f) FeSb$_2$: the spin rotates in-plane by an azimuthal angle (dashed arc) toward the energetically favored direction (light-blue arrow).}
    \label{fig:fig2}
\end{figure*}

In the following, we are going to demonstrate an \textit{altermagnet molecule}, in which the out-of-plane moment survives down to an isolated magnetic site, with spin canting spontaneously off the primary axis because of SOC. In Fig. \ref{fig:fig2}a, we employ an octahedral crystal field model but rotate the octahedron to a trigonal configuration, which mimics the MnTe$_6$ local atomic structure in the MnTe crystal, as shown in Fig. \ref{fig:fig2}b. 
Under a trigonal field, the $O_h$ point group reduces to $D_{3d}$, which is precisely the site symmetry of Mn atoms in MnTe. Then the $t_{2g}-e_g$ crystal field levels transform to $a_{1g}\oplus e_g^{\pi} - e_g^{\sigma}$ in the $D_{3d}$ representation. 
Since $a_{1g}$ is a singlet with $l_z=0$ and $e_g^{\pi}$ is a doublet with $\langle l_z\rangle=\pm1$, the $t_{2g}$ triplet carries no net orbital moment; together with the $e_g^{\sigma}$ doublet the total $l_z$ vanishes for the $d^5$ high-spin ($^6S$, $L=0$) configuration of MnTe in the absence of SOC.

For convenience, we adopt the $|l_z\rangle$ basis ordered as $(|{+}2\rangle,|{+}1\rangle,|0\rangle,|{-}1\rangle,|{-}2\rangle)$ for the crystal field Hamiltonian. We consider that the exchange field or Coulomb interaction ($\Delta_{ex}$) is far larger than the crystal field splitting ($\Delta_{cf}$) to keep the high-spin state. Then we introduce atomic SOC to this molecular model in the following Hamiltonian under the $O_h$ point group, 
\begin{equation}
H = H_{cf} - \Delta_{ex} {\bm J}\!\cdot\!\bm\sigma
+ \lambda {\bm L\!\cdot\!\bm \sigma},
\label{eq:Hloc}
\end{equation}
where $\bm L$ and $\bm \sigma$ orbital operator and spin Pauli matrices, respectively, $\bm J = (\cos\phi,\sin\phi,0)$ represent the in-plane spin direction. $H_{cf}$ is the octahedral crystal field,
\begin{equation}
H_{cf}^A \;=\; \frac{\Delta_{cf}}{3}\!
\begin{pmatrix}
1 & 0 & 0 & \sqrt{2} & 0 \\
0 & 2 & 0 & 0 & -\sqrt{2} \\
0 & 0 & 0 & 0 & 0 \\
\sqrt{2} & 0 & 0 & 2 & 0 \\
0 & -\sqrt{2} & 0 & 0 & 1
\end{pmatrix},
\label{eq:HCF_A}
\end{equation}
where only the $\delta l_z =\pm 3$ off-diagonals are allowed by the three-fold rotational symmetry. Here, the superscript $A$ represents the sublattice. For the second sublattice $B$ that rotates by 180$^\circ$ from $A$ (see Fig. 2b), these off-diagonals reverse sign in Eq.~\ref{eq:HCF_A}. $H_{cf}^{A,B}$ diagonalizes to $\{0,0,0,\Delta_{cf},\Delta_{cf}\}$, namely the $a_{1g}\oplus e_g^{\pi}$ triplet at $0$ and the $e_g^{\sigma}$ doublet at $\Delta_{cf}$. In addition, a trigonal distortion can further split the  $a_{1g}\oplus e_g^{\pi}$ triplet. As we will show that spin canting exists already in the ideal octahedral field. The additional trigonal distortion field $H_{\rm tri}=\frac{\delta_{\rm tri}}{6}\bigl(3L_z^{2}-6\bigr)$ is diagonal in the $|l_z\rangle$ basis and still preserves the local $C_{3v}$ symmetry, so it only modifies the spin canting amplitude in a perturbative manner (see Sec. 2 of SI). Thus, we omit the trigonal distortion for simplicity hereafter.

The large in-plane exchange coupling ($\Delta_{ex}$) splits the crystal field levels into two spin channels while the out-of-plane orbital and spin moments remain zero. Only after including SOC, two spin channels get mixed to induce net out-of-plane magnetization (see Fig. 2a). Diagonalization of Eq.~\ref{eq:Hloc} at the in-plane spin angle $\phi$ (see details in SI) followed by the $d^5$ occupied-state trace
$ M_z(\phi) \;=\; \sum_{n\,\in\,\mathrm{occ}}\langle n|L_z|n\rangle $
yields a strict third harmonic,
\begin{equation}
M_z(\phi) \;=\; A_3\sin(3\phi),
\label{eq:Mz_third}
\end{equation}
with the prefactor obeying the leading-order perturbative scaling confirmed numerically,
\begin{equation}
A_3 \;\propto\; \lambda^{3}\,\Delta_{cf}^{\,3}\,\Delta_{ex}^{-6}.
\label{eq:scaling}
\end{equation}
The $\lambda^{3}$ onset is the lowest order at which the symmetric cancellation of the $d^{5}$ ground state is overcome: zeroth, first, and second order in SOC vanish after the occupied-state sum, so the moment first appears at third order. The remaining dependence, $\Delta_{cf}^{3}\,\Delta_{ex}^{-6}$, is obtained numerically from exact diagonalization while the $\sin(3\phi)$ form is fixed by the threefold rotational symmetry. Experimentally, the $\sin(3\phi)$ dependence is supported by the field-driven reduction of six in-plane domains to three observed by X-ray magnetic linear/circular dichroism \cite{Amin2024}, whereas resolving the full angular profile requires fine-controlled rotation of the Néel vector.
In addition, the $\lambda^{3}$ indicates a sensitive SOC dependence of the residual moment and AHE, which may be demonstrated through chemical doping, for example, in Mn(Te$_{1-x}$Se$_{x}$).

Between $A$ and $B$ sublattices, the crystal field and exchange direction rotate under the $C_{2z}$ rotation while $M_z$ is invariant since it is the $z$ component of an axial vector. Each sublattice carries its own $D_{3d}$ site group, and $C_{2z}$ maps one onto the other, so that the two conjugate $D_{3d}$ groups together generate the $D_{6h}$ point group of the lattice model (Eq.~\ref{eq:H}). This is consistent with the observation from Eqs.~\ref{eq:Mz_third} and~\ref{eq:scaling}, where two sublattices exhibit the same orbital magnetization after reversing the spin angle and crystal field together since both $\sin(3\phi)$ and $\Delta_{cf}$ reverse their signs, 
\begin{equation}
M_z^A(\phi) \;=\; M_z^B(-\phi). 
\end{equation}
The molecule model therefore produces a finite, sublattice-even orbital magnetization, which is consistent with the anomalous Hall response of the continuum lattice model. Different from an ordinary antiferromagnet where two spin sublattices share the same local crystal field, the altermagnet hosts two spin sublattices with opposite crystal fields. Such a crystal field - sublattice locking leads to unique properties, including the spontaneous net magnetization for AHE altermagnets \cite{zhao2026prb}.  

Because of SOC, orbital magnetization $M_z$ comes together with an out-of-plane spin component $S_z$.
Such a spin canting follows the same $\sin{(3\phi)}$ dependence as $M_z$. We can easily rationalize $S_z$ from the octahedral atomic structure. At $\phi = 0^\circ$ (green arrow in Fig. 2b), the in-plane spin points to the middle point of the octahedra edge (Te-Te bond). The spin canting up and down makes no difference in energy and therefore the spin remains in-plane. In contrast, at $\phi = 30^\circ$ (red arrow in Fig. 2b) or $-30^\circ$ (blue arrow in Fig. 2b), the spin points to the octahedral corner (a Te atom) when canting up, while it points to the edge (Te-Te bond) when canting down, resulting in different energies. Consequently, the spin will spontaneously choose an energetically favored canting direction. 
In the molecular-field model, magnetic anisotropy \cite{Moriya1960} itself induces spin canting and weak net magnetization due to SOC.   

The altermagnet molecule model that rationalizes the spin canting by crystal field symmetry can be readily extended to other altermagnet candidates. In another g-wave altermagnet CrSb (Fig.~2c), the spin moment points along the $z$ axis (dashed line), i.e., toward the center of an octahedral face. Because this face is an equilateral triangle, the three possible horizontal tilts are equivalent, and no canting is favored. In RuO$_2$ and KV$_2$Se$_2$O (d-wave altermagnets in Fig.~2d and 2e), the spin moment points toward the midpoint of an octahedral edge, so tilting toward either side of the edge costs the same energy. In both cases, the moment stays along its primary axis and two sublattices cancel exactly, so neither a net orbital moment nor an anomalous Hall effect emerges. By contrast, in FeSb$_2$ (d-wave in Fig.~2f), the spin moment lies in-plane and the octahedron is vertically strained, so the triangle that the spin faces is no longer equilateral. Canting up or down remains degenerate, but canting horizontally toward a ligand atom or toward an edge (dashed arc) now costs different energies. The dark-blue spin therefore rotates by a finite azimuthal angle to the light-blue spin to lower its energy. Such a horizontal canting is favored in the same direction for two sublattices, thereby giving rise to a net magnetic moment and an accompanying anomalous Hall effect.

The local molecular model and the lattice model are complementary rather than redundant. The octahedral molecule gives the real-space, atomic picture: it shows intuitively how SOC tilts the spin out of plane ($S_z$) and generates an on-site orbital moment ($M_z$), arising solely from the interplay of the local crystal field and SOC without invoking any exchange between neighboring spins. $M_z$ reproduces the nonlinear SOC dependence found in the lattice model. This local orbital moment corresponds to the wave-packet self-rotation -- essentially the trivial
part $M_z^{\rm tri}$ of the orbital magnetization -- and does not by itself
produce a Hall response. When forming an extended lattice, inter-site hopping between these molecular levels gives rise to an extra orbital component -- the topological part $M_z^{\rm topo}$ which originates
from the mass-center circulation of the wave packet. Thus, for transport related to the global charge motion, only the topological orbital component contributes to the Hall conductivity through the
St\v{r}eda relation.

\section{Discussion}
Our results indicate that the weak out-of-plane ferromagnetism is a ground state property of the AHE altermagnet\cite{liu2026observation}. 
Despite its small values, the chemical potential gradient of orbital magnetization governs the AHC via the St\v{r}eda relation. In the $g$-wave altermagnet, because of the energetically favored $\sin(3\phi)$ angle-dependence of spin canting, an external magnetic field can flip the out-of-plane magnetization and spontaneously rotate the spin by $\pm 60^\circ$ in-plane, giving rise to the field control of the N\'{e}el vector in a hysteretic way. Therefore, the remanent ferromagnetization is essential to understand the N\'{e}el order switching and anomalous Hall response for unconventional antiferromagnets. 

\section{Methods}
\subsection{Tight-binding model}
The tight-binding model is constructed based on the magnetic atom Wyckoff site symmetry and the space group symmetry in Ref.~\cite{Roig2024PRB}. The terms in Eq. \ref{eq:H} are as follows,
\begin{align*}
\varepsilon_{0,\bm k} &= t_1\!\left[\cos k_x + 2\cos\!\tfrac{k_x}{2}\cos\!\tfrac{\sqrt{3}k_y}{2}\right]
                       + t_2\cos k_z - \mu, \\
t_{x,\bm k} &= t_3\cos\!\tfrac{k_z}{2}, \\
t_{z,\bm k} &= t_4\,\sin k_z\;f_y\,(f_y^2 - 3f_x^2), \\
\lambda_{x,\bm k} &= \lambda\cos\!\tfrac{k_z}{2}\,(f_x^2 - f_y^2), \\
\lambda_{y,\bm k} &= -\,2\lambda\cos\!\tfrac{k_z}{2}\,f_x f_y, \\
\lambda_{z,\bm k} &= \lambda \sin\!\tfrac{k_z}{2}\,f_x\,(f_x^2 - 3f_y^2),
\end{align*}
with
\begin{align*}
f_{x,\bm k} =& \sin k_x + \sin\!\tfrac{k_x}{2}\cos\!\tfrac{\sqrt{3}k_y}{2}, \\
f_{y,\bm k} =& \sqrt{3}\,\cos\!\tfrac{k_x}{2}\sin\!\tfrac{\sqrt{3}k_y}{2}.
\end{align*}

The Brillouin-zone integrals of AHC and orbital magnetization are evaluated on a uniform $200\times200\times200$ mesh. Other details can be found in the SI.

\section{Data availability}
All study data are included in the article.

\section{Acknowledgments}
We thank helpful discussions with Igor I. Mazin and Kirill D. Belashchenko. 

\section{Author contributions}
B.Y. conceived the project. Y.Z. and B.Y. performed calculations and wrote the manuscript with input from all authors. Y.J., K.D. and C.X.L. reviewed and analyzed the results. All the authors approved the manuscript.

\section{Funding}
B.Y., C.X.L, and Y.J. acknowledges the financial support by the Penn State Materials Research Science and Engineering Center for Nanoscale Science (MRSEC) under National Science Foundation (NSF) award DMR-2011839.

\section{Competing interests}
The authors declare no competing interests.

\renewcommand{\bibsection}{\section*{References}}

%

\end{document}